\documentclass[%
reprint,
amsmath,
amssymb,
floatfix,
aip,
cha,
english,
]{revtex4-1}

\usepackage{graphicx}
\usepackage{dcolumn}
\usepackage{bm}
\usepackage{hyperref}
\usepackage{color}

\begin{document}

\preprint{APS/123-QED}

\title{A variable partially-polarizing beam splitter}

\author{Jefferson Fl\'orez}
\email{jflor020@uottawa.ca}


\author{Nathan J. Carlson}

\author{Codey H. Nacke}

\author{Lambert Giner}

\author{Jeff S. Lundeen}

\affiliation{Department of Physics and Centre for Research in Photonics, University
of Ottawa, 25 Templeton Street, Ottawa, Ontario K1N 6N5, Canada}

\date{\today}

\begin{abstract}
We present designs for variably polarizing beam splitters. These are
beam splitters allowing the complete and independent control of the
horizontal and vertical polarization splitting ratios. They have quantum
optics and quantum information applications, such as quantum logic
gates for quantum computing and non-local measurements for quantum
state estimation. At the heart of each design is an interferometer.
We experimentally demonstrate one particular implementation, a displaced
Sagnac interferometer configuration, that provides an inherent instability
to air currents and vibrations. Furthermore, this design does not require any custom-made optics but only common components which can be easily found in an optics laboratory.
\end{abstract}

\maketitle

\section{\label{Intro}Introduction}

\begin{figure}[htbp]
\centering \includegraphics[width=\linewidth]{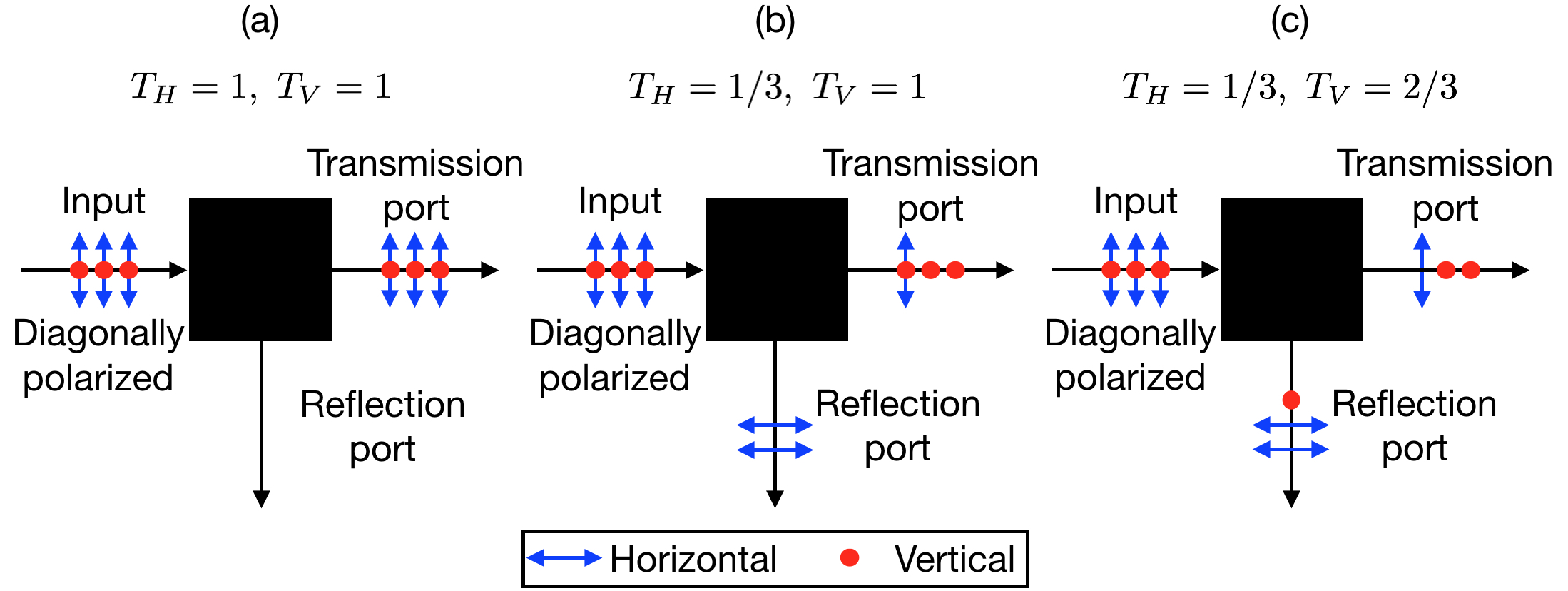} \caption{Schematic representation of a partially-polarizing beam splitter.
If $T_{H}=1=T_{V}$ as in panel (a), one will have both the horizontally
and vertically polarized light being transmitted by the PPBS, whereas
there will be no light coming out at the reflection port. If $T_{H}=1/3$
and $T_{V}=1$ as in panel (b), one will have one third of the incoming
horizontally polarized light being transmitted and the rest reflected,
whereas all the vertically polarized light is transmitted by the PPBS.
Finally, if $T_{H}=1/3$ and $T_{V}=2/3$ as in panel (c), one third
and two thirds of the horizontally and vertically polarized light
are transmitted, respectively, and the rest is reflected.}
\label{Sketch} 
\end{figure}

Typical polarizing beam splitters are intended to spatially separate
the horizontal ($H$) and vertical ($V$) polarization components
of an input beam. However, there are several applications in which
a particular set of transmission and reflection coefficients for each
polarization are required, like in quantum logic gates \cite{Kiesel05,Okamoto05,Langford05},
quantum state estimation techniques \cite{Ling06,Medendorp11}, and
wave-particle duality studies \cite{Kaiser12}. A device that provides
such coefficients is called a partially-polarizing, or polarization-dependent,
beam splitter (PPBS). To illustrate its properties, consider a PPBS
illuminated by a diagonally polarized beam, as shown in Fig. \ref{Sketch}.
Depending on the values of the transmission ($T$) and reflection
($R$) coefficients for $H$ and $V$, one can have any chosen splitting
ratio (i.e., $T$:$R$) between the two PPBS output ports independently for the horizontally
and vertically polarized light.

A PPBS can be built using multilayered dielectric coatings designed
for specific $T$ and $R$ coefficients for each polarization. The
drawback of this custom fabricated beam splitter is that such coefficients
cannot be tuned for different purposes, including the correction of
fabrication caused deviations from the target values of $T$ and $R$.
These can lead to a reduced performance in some applications \cite{Kaiser12}.
Moreover, these devices are not available off-the-shelf and are, thus,
expensive. In this paper, a \emph{variable} partially-polarizing beam
splitter (VPPBS) is introduced featuring a complete and independent
control of the horizontal and vertical $T$ and $R$ coefficients.
Furthermore, it is based on bulk optical components that are usually
available in any optics laboratory.

\begin{figure}[htbp]
\centering \includegraphics[width=\linewidth]{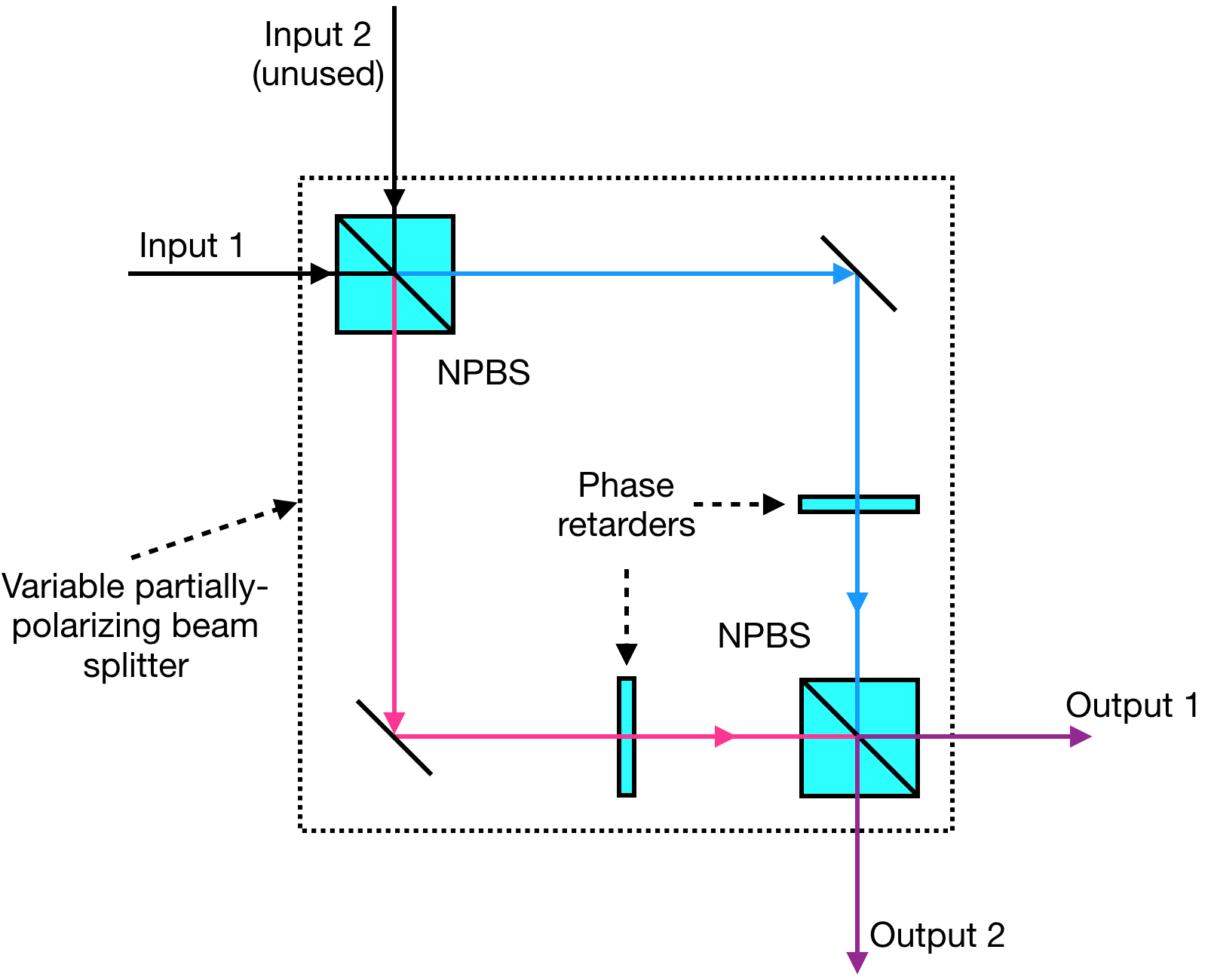} \caption{The basis of the variable partially-polarizing beam splitter (VPPBS)
presented here is two independent interference processes occurring
in a Mach-Zehnder interferometer, one for the horizontal and the other
for the vertical polarization. The phase retarders introduce polarization-dependent
phases between the two optical paths. Outputs 1 and 2 play the role
of the transmission and reflection VPPBS ports, respectively. NPBS
is a non-polarizing 50:50 beam splitter. That is, a beam-splitter
with $T=R=1/2$ for both polarizations.}
\label{MZ} 
\end{figure}

The working principle of the VPPBS presented here is the interference
of two beams in a simple interferometer like a Mach-Zehnder. Consider
light entering only one input of the first beam splitter in that interferometer,
as shown in Fig. \ref{MZ}. Light interferes constructively or destructively
at the last beam splitter depending on the phase between the two optical
paths. It follows that the interferometer input light can be made
to exit the last beam splitter entirely via Output 1 or, alternately,
entirely via Output 2, or some combination of the two output ports.
By tuning the interferometer phase one can set any desired splitting
ratio between these two output ports. Considered in its entirety,
the Mach-Zehnder is a beampslitter, with Output 1 and 2 arbitrarily
defined as the effective transmission and reflection beampslitter
ports. And so, one can vary the beam splitter $T$ and $R$ coefficients
by varying the phase. Now consider both $H$ and $V$ entering the
interferometer input. By tuning the phase for each of these polarizations
independently the splitting ratios for $H$ and $V$ can be set independently.
With this independent phase control, the Mach-Zehnder is a VPPBS.

\section{Theory}

\label{Theory} In this section a theoretical description of the VPPBS
mechanism outlined above is presented. (For completeness, in Appendix \ref{OtherVPPBSs}, a theoretical description of a nominally distinct,
but actually closely related, configuration is presented based on polarizing
beam splitters in the place of non-polarizing beam splitters.) Consider an incoming light beam
entering at Input 1 of the Mach-Zehnder interferometer in Fig. \ref{MZ}.
In the $H/V$ basis, such a light beam is characterized by an electric
field $\mathbf{E}_{\text{in}}(t)$ of the form 
\begin{equation}
\mathbf{E}_{\text{in}}(t)=\left[\begin{array}{c}
E_{\text{in}}^{H}(t)\\
\\
E_{\text{in}}^{V}(t)
\end{array}\right],
\label{Ein}
\end{equation}
where $E_{\text{in}}^{H}(t)$ and $E_{\text{in}}^{V}(t)$ are the
$H$ and $V$ polarization components, respectively. After the first
50:50 non-polarizing beam splitter (NPBS) in Fig. \ref{MZ}, the electric
fields describing the upper and lower paths just before the phase
retarders differ by a phase of $\pi/2$ due to the different number
of reflections. This is, 
\begin{equation}
\mathbf{E}_{\text{upper}}(t)=\cfrac{1}{\sqrt{2}}\mathbf{E}_{\text{in}}(t),\quad\mathbf{E}_{\text{lower}}(t)=\cfrac{e^{i\pi/2}}{\sqrt{2}}\mathbf{E}_{\text{in}}(t).
\end{equation}

Now, the two phase retarders in the interferometer arms introduce
independent phases $\phi_{H}$ and $\phi_{V}$ to the $H$ and $V$
polarization, respectively. These phase retarders can be, for example,
two liquid crystal cells with crystal axes orthogonally oriented to
one another, i.e., along either the $H$ or $V$ directions. The electric
fields just after the phase retarders and before the second NPBS in
Fig. \ref{MZ} read 
\begin{gather}
\tilde{\mathbf{E}}_{\text{upper}}(t)=\cfrac{1}{\sqrt{2}}\left[\begin{array}{c}
E_{\text{in}}^{H}(t)\\
\\
e^{i\phi_{V}}E_{\text{in}}^{V}(t)
\end{array}\right],\\
\nonumber \\
\tilde{\mathbf{E}}_{\text{lower}}(t)=\frac{e^{i\pi/2}}{\sqrt{2}}\left[\begin{array}{c}
e^{i\phi_{H}}E_{\text{in}}^{H}(t)\\
\\
E_{\text{in}}^{V}(t)
\end{array}\right],
\end{gather}
where it has been assumed without loss of generality that the phase
retarder in the upper path introduces the phase $\phi_{V}$, while
the one in the lower path introduces $\phi_{H}$. After the second
NPBS, i.e. at the outputs of the Mach-Zehnder interferometer, the
electric fields become 
\begin{gather}
\mathbf{E}_{\text{out,1}}(t)=\left[\begin{array}{c}
e^{i\phi_{H}/2}E_{\text{in}}^{H}(t)\cos\left(\cfrac{\phi_{H}}{2}\right)\\
\\
e^{i\phi_{V}/2}E_{\text{in}}^{V}(t)\cos\left(\cfrac{\phi_{V}}{2}\right)
\end{array}\right],\label{Eout1}\\
\nonumber \\
\mathbf{E}_{\text{out,2}}(t)=\left[\begin{array}{c}
-ie^{i\phi_{H}/2}E_{\text{in}}^{H}(t)\sin\left(\cfrac{\phi_{H}}{2}\right)\\
\\
ie^{i\phi_{V}/2}E_{\text{in}}^{V}(t)\sin\left(\cfrac{\phi_{V}}{2}\right)
\end{array}\right].\label{Eout2}
\end{gather}

The $T$ coefficient for the horizontal polarization is defined as
the ratio between the horizontal light intensity in Output 1 and the
intensity of that polarization in the input beam. Similarly, $T_{V}$
corresponds to the ratio between the vertical light intensity in Output
1 and the initial intensity of such polarization. In terms of Eqs.
(\ref{Eout1}) and (\ref{Eout2}), this is 
\begin{gather}
T_{\epsilon}\equiv\frac{|E_{\text{out,1}}^{\epsilon}(t)|^{2}}{|E_{\text{in}}^{\epsilon}(t)|^{2}}=\frac{1+\cos\phi_\epsilon}{2},\label{TH}
\end{gather}
where $\epsilon=H,V.$ The fields $E_{\text{out,1}}^{H}$ and $E_{\text{out,1}}^{V}$
are, respectively, the $H$ and $V$ components of $\mathbf{E}_{\text{out,1}}$.
The reflection coefficients for the horizontal and vertical polarizations
are given by $R_{H}=1-T_{H}$ and $R_{V}=1-T_{V}$, respectively.
Thus, by tuning $\phi_{H}$ and $\phi_{V}$ any possible value of
reflection and transmission coefficients can be chosen. This constitutes
a VPPBS.

The challenge in this design is how to vary $\phi_{H}$ and $\phi_{V}$. 
As mentioned earlier, one possibility for the phase retarders
are orthogonally oriented two liquid crystals. Here, the relative
phases $\phi_{H}$ and $\phi_{V}$ are independently tuned by means
of the AC voltage applied to each liquid crystal. Another possibility,
which is the one implemented here, is to vary the tilt of a uniaxial
birefringent crystal (i.e., one with parallel input and output faces)
in one of the interferometer arms. Both the refractive index and optical
path length will vary differently for the two polarizations as the
element is tilted. In turn, the introduced phases $\phi_{b}^{H}$
and $\phi_{b}^{V}$ for the the $H$ and $V$ polarizations, respectively,
will be differently tuned by the tilt. This will set the phase difference
$\Delta\phi=\phi_{H}-\phi_{V}$. To achieve full independent control
of each phase, tilting a second non-birefringent plate (e.g., glass)
can be used to introduce an identical phase $\phi_{g}$ offset to
the two polarizations, so that 
\begin{gather}
\phi_{H}=\phi_{b}^{H}-\phi_{g},\label{phiH}\\
\phi_{V}=\phi_{b}^{V}-\phi_{g}.\label{phiV}
\end{gather}

In summary, with the two tilts as control parameters, it is possible
to independently set the two system degrees of freedom, $\phi_{H}$
and $\phi_{V}$.

The relative phase $\phi$ introduced between the two optical paths
in a Mach-Zehnder interferometer by a plate of thickness $d$ and
refractive index $n$ is $\phi=2\pi d(n-n_{a})/\lambda$, where $n_{a}$
is the refractive index of air and $\lambda$ is the wavelength of
light. This expression describes the case when the plate is placed
in one arm of the interferometer so that it is normal the beam. In
Appendix \ref{TiltMedia}, it is shown that if the plate is tilted by
$\theta_{p}$ from normal, the relative phase is given by
\begin{equation}
\phi_{p}^{\epsilon}=\frac{2\pi d_{p}}{\lambda}\left[\frac{n_{p}^{\epsilon}}{\cos\theta'^{\epsilon}_{p}}-n_{a}\left(\cos\theta_{p}+\sin\theta_{p}\tan\theta'^{\epsilon}_{p}\right)\right],\label{phi}
\end{equation}
where $p=b$ or $g$ for the birefringent and glass plates, respectively, and $\theta'^\epsilon_p$ is the angle of light inside the optical medium after refraction. 
In the case, $p=g$ the $\epsilon$ label is not used throughout the
paper, whereas for $p=b$, $\epsilon=H,V$. In particular, there are
two refractive indices $n_{b}^{H}$ and $n_{b}^{V}$ for the birefrigent
medium. The angle $\theta'^\epsilon_p$
is given by Snell's law, 
\begin{equation}
\theta'^{\epsilon}_{p}=\arcsin\left(\frac{n_{a}}{n_{p}^{\epsilon}}\sin\theta_{p}\right).\label{Snell}
\end{equation}
For simplicity, $\theta'_{b}\equiv \theta'^{H}_{b}$ in the reminder of the paper.

In the present work, the birefringent crystal is tilted around the laboratory
vertical axis, which is parallel to the vertical polarization and
perpendicular to the optical table, as shown in Fig. \ref{ExpSetup}.
Furthermore, the optic axis $\vec{c}$ of the crystal lies on the
horizontal plane, which means that the angle $\theta_{c}'$ between
$\vec{c}$ and the beam propagation direction changes as the crystal
is allowed to rotate around the vertical axis. In this arrangement,
the refractive index is constant for the vertical (ordinary) polarization,
whereas the effective refractive index for the horizontal (extraordinary)
polarization changes as the crystal is tilted according to \cite{Boyd},
\begin{equation}
n_{b}^{H}=\left[\frac{\cos^{2}(\theta_{c}')}{n_{o}^{2}}+\frac{\sin^{2}(\theta_{c}')}{n_{e}^{2}}\right]^{-\frac{1}{2}},\label{nbH}
\end{equation}
with $n_{o}$ and $n_{e}$ the ordinary and extraordinary refractive
indices of the birefringent crystal, respectively, and $\theta_{c}'$
defined as $\theta_{c}'=\theta_{c}+\theta_{b}'$. We allow for the fact that
the chosen crystal might potentially be cut so that crystal axis $\vec{c}$
is at an angle $\theta_{c}$ to the crystal face normal. For the vertical
polarization, $n_{b}^{V}$ is identically equal to $n_{o}$, whereas
for the horizontal polarization the refractive index $n_{b}^{H}$
depends on the crystal tilt, as seen from the last equation.

According to Eq. (\ref{nbH}), one must know $\theta_{b}'$ to find
$n_{b}^{H}$, but at the same time one needs $n_{b}^{H}$ to get $\theta_{b}'$
by means of Eq. (\ref{Snell}). Unfortunately, this pair of equations
do not have analytic solution for $\theta_{b}'$ and $n_{b}^{H}$.
So, in order to be able to contrast the experimental results in Sec. \ref{ExpResults} with the theoretical predictions, $n_{b}^{H}$ is estimated
in the following way. First, $\theta_{b}'$ is approximated by $\theta_{b}$
in Eq. (\ref{nbH}), resulting in $\theta_{c}'\approx\theta_{c}+\theta_{b}$
which is used to find a zero-order approximation for $n_{b}^{H}$.
Second, using this result, a value for $\theta_{b}'$ is calculated
via Eq. (\ref{Snell}). Third, that result for $\theta_{b}'$ is substituted
in Eq. (\ref{nbH}) to finally obtain a first-order approximation
for $n_{b}^{H}$. Repeating the same steps, a second-order approximation
for $n_{b}^{H}$ can be calculated. The discrepancy between the zero-
and first-order approximations is less than 0.5\% and between the
first- and second-order approximations is less than 0.004\%. In this
paper, the second-order approximation is used.

\section{\label{ExpRealization}Experimental realization}

\begin{figure}[htbp]
\centering \includegraphics[width=\linewidth]{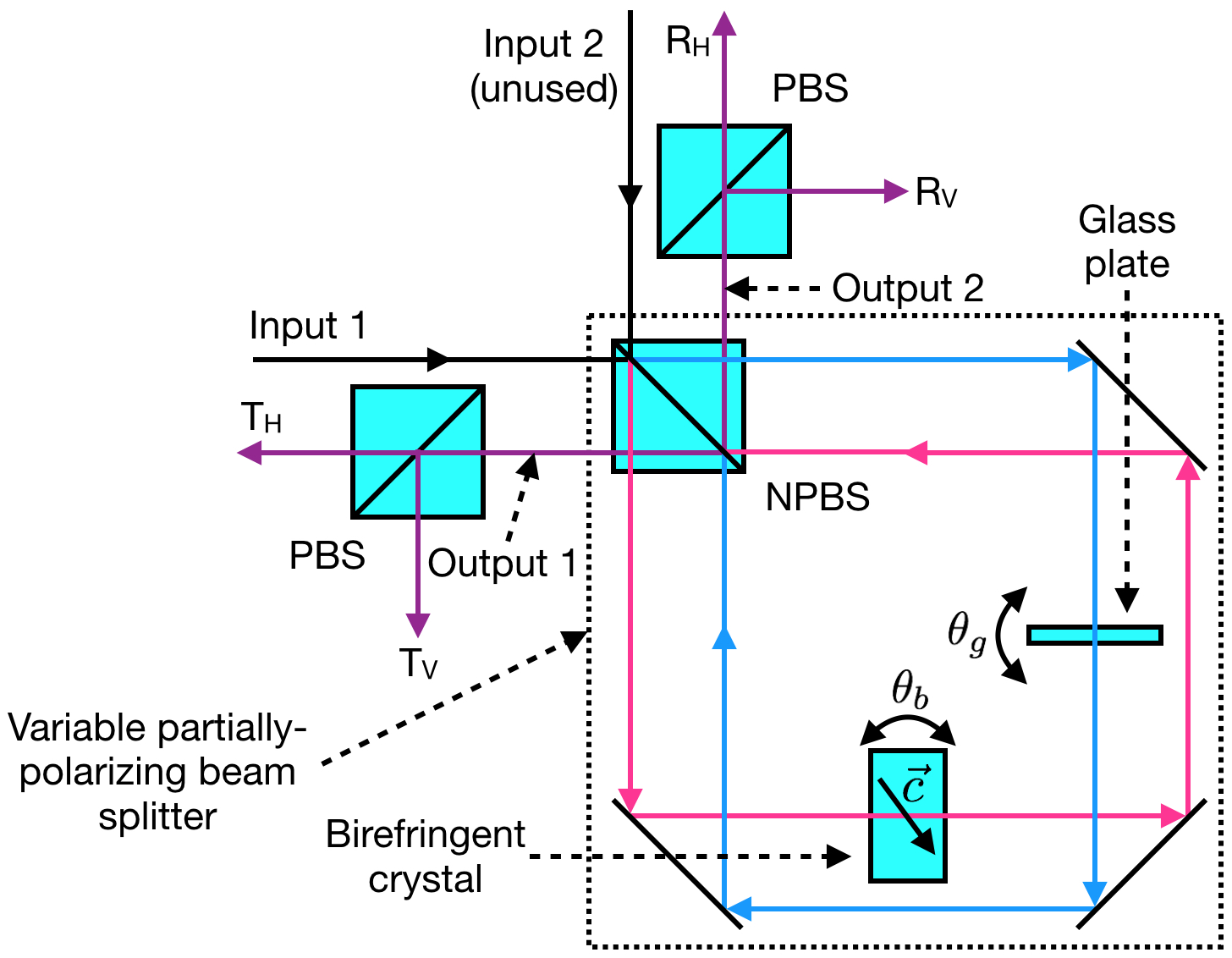} \caption{Experimental realization of a variable partially-polarizing beam splitter
using a displaced Sagnac interferometer composed of three mirrors
and a non-polarizing beam splitter (NPBS). Two polarizing beam splitters
(PBSs) have been placed at each output port to study the transmission
and reflection coefficients for each polarization. The thickness of
the birefringent crystal has been exaggerated to indicate the orientation
of its optic axis $\vec{c}$.}
\label{ExpSetup} 
\end{figure}

The Mach-Zehnder interferometer shown in Fig. \ref{MZ} can be used
to implement a VPPBS. However, depending on its spatial dimensions
and external factors like vibrations and air currents, this interferometer
might require active phase stabilization in order to hold a specific
set of $T$ and $R$ coefficients for a long period of time, e.g.
hours. A variation of
the Mach-Zehnder interferometer, called a displaced Sagnac interferometer\cite{Nagata07,Micuda14,Ashby16}, is used instead to reduce this inherent instability.
In this, the light is split by and returns to the same NPBS using three mirrors,
as shown in Fig. \ref{ExpSetup}. The two counter-propagating beams
inside the interferometer play the same role as the two arms in a
Mach-Zehnder. In the non-displaced version of the Sagnac interferometer,
the counter-propagating beams inside the interferometer follow exactly
the same paths. This makes it difficult to introduce a relative phase
between the beams, as it is required in the current scheme. It also
means that beam in Output 1 exits along the exact path of the input
beam, which makes the output beam difficult to access.  A displaced Sagnac eliminates these issues. In it, one translates
the mirror in the Sagnac that is diagonally opposite to the NPBS.
This separates the two counter-propagating paths, while maintaining
their collinearity. 

Given that the transverse separation between the paths is small ($\sim2$
cm) compared to the footprint of the interferometer (70 cm$\times$
70 cm), and the fact that the beams are reflected and transmitted
by the same mirrors and NPBS in the interferometer, any vibration
or air current affects both optical paths roughly in the same way,
making the Sagnac interferometer stable without active stabilization\cite{Micuda14}.

The experiment was carried out using a HeNe laser in free space with
wavelength $\lambda=632.8$ nm. The displaced Sagnac interferometer
was built using a broadband (400 - 700 nm) 50:50 NPBS cube with 2.54
cm side length (Thorlabs BS013) and three 5.08 cm diameter silver
mirrors. As mentioned above, the phase retarders can be either two
liquid crystals driven by an AC voltage or a birefringent crystal
plus a glass plate tilted as shown in Fig. \ref{ExpSetup}. However,
only the second case was investigated here since such elements are easily
accessible in the laboratory. For a full comparison between the tilted optical medium and liquid crystal methods we refer the reader to the Table \ref{comparison}. The birefringent crystal was a $\beta$ barium
borate (BBO) crystal of nominal thickness $d_{b}=0.245$ mm and nominal optic
axis at $\theta_{c}=33.4^{\circ}$; the glass plate was a microscope cover slide of nominal thickness $d_{g}=0.16$ mm. These two elements were mounted in
automated rotation stages and tilted between $-10^{\circ}$ and $10^{\circ}$
in steps of $0.1^{\circ}$ for $\theta_{b}$ and $0.25^{\circ}$ for
$\theta_{g}$. The input light was diagonally polarized by means of
a polarizing beam splitter (PBS) plus a half-wave plate with its fast
axis at $22.5^{\circ}$ with respect to the horizontal direction.
As seen in Fig. \ref{ExpSetup}, at each output port of the VPPBS
a PBS was placed to study the $T$ and $R$ coefficient for both polarizations.
The intensities were recorded for each value of $\theta_{b}$ and
$\theta_{g}$ by four photodiodes after averaging 100 measurements
taken over 2 seconds.
\onecolumngrid
\begin{center}
\begin{table}
\caption{\label{comparison}{Comparison between tilted optical medium and liquid crystal methods to control the relative phase in the displaced Sagnac interferometer in Fig. \ref{ExpSetup}.}}
\begin{ruledtabular}
\begin{tabular}{l  l  l}
  & Tilted birefringent crystal and glass plate & Liquid crystals\\\hline
  Advantages & Need to align axis only for the birefringent crystal & $\phi_H$ and $\phi_V$ each have their own control\\
  \\
  & Induced phase is stable over months & Normal incidence (i.e. no beam displacement),\\ 
  & & implying bigger aperture for a given device width\\
  \\
  & Tilted medium is usable to its edge, which allows & \\
  & for counterpropagating beam clearance & \\\hline
  Disadvantages & $\phi_H$ and $\phi_V$ are not independently controlled & Need to align axes for both liquid crystals\\
  \\
  & Refractive beam displacement can reduce & Liquid crystal response can change from one\\
  & interference visibility & voltage ramp to the next\\
  \\
  & Requieres a precision rotation mount (e.g. vernier) & Since they are not usable up to their edge, a large\\
  & & Sagnac path separation is required\\
  \\
  & & Easy to damage with a DC voltage\\
\end{tabular}
\end{ruledtabular}
\end{table}
\par\end{center}
\twocolumngrid

\section{\label{ExpResults}Experimental results}

To demonstrate the working principle of a VPPBS based on the Sagnac
interferometer in Fig \ref{ExpSetup}, the four coefficients $T_{H}$,
$T_{V}$, $R_{H}$ and $R_{V}$ were measured as described in Sec. \ref{ExpRealization} and are shown in Fig. \ref{data1}(a).

\onecolumngrid 
\begin{center}
\begin{figure}[h!]
\centering \includegraphics[width=0.497\linewidth]{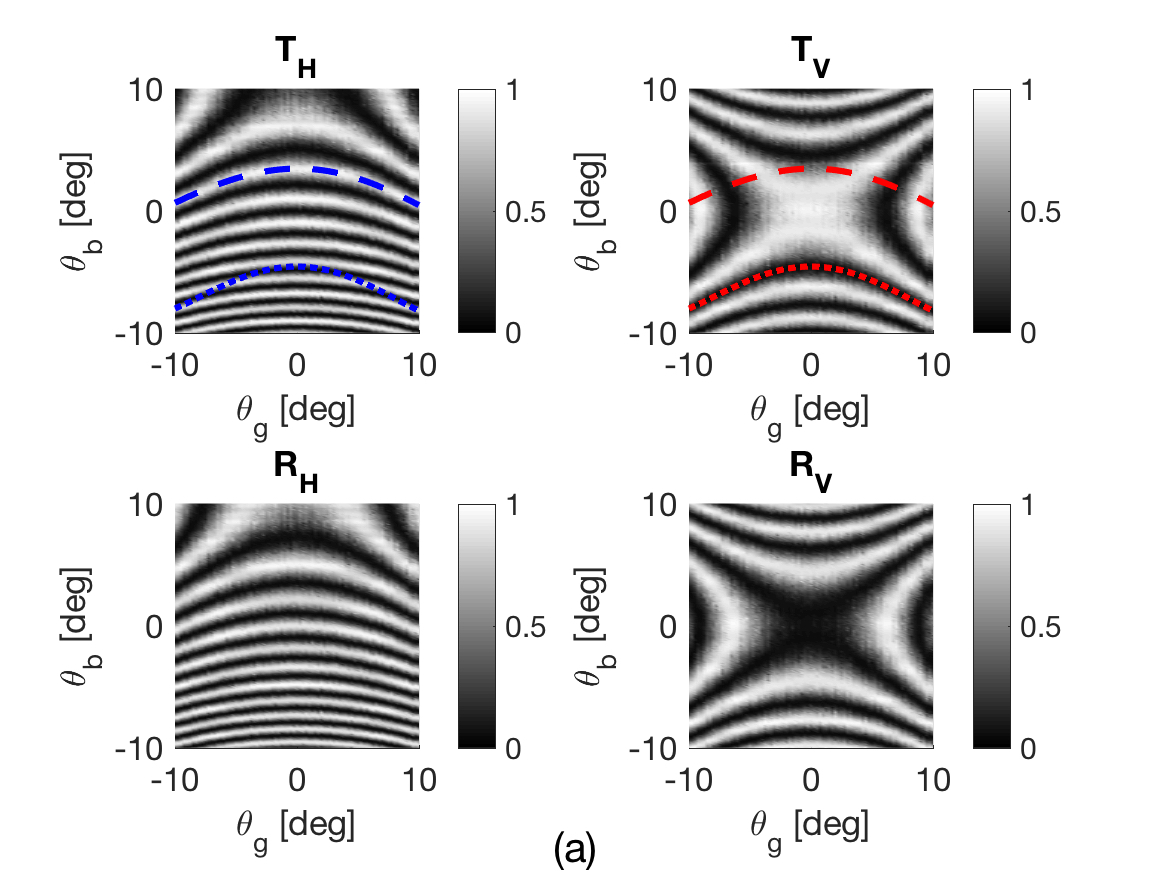} \centering
\includegraphics[width=0.497\linewidth]{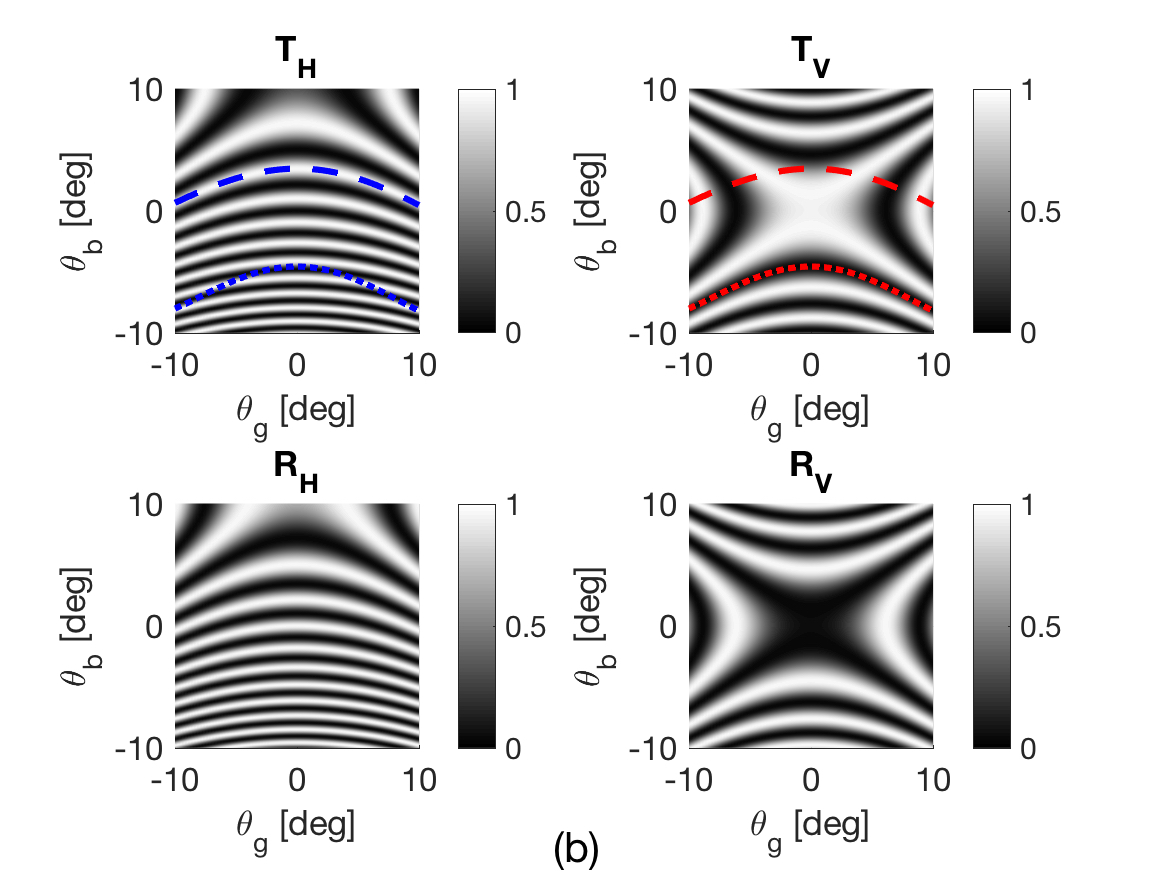} 
\includegraphics[width=0.497\linewidth]{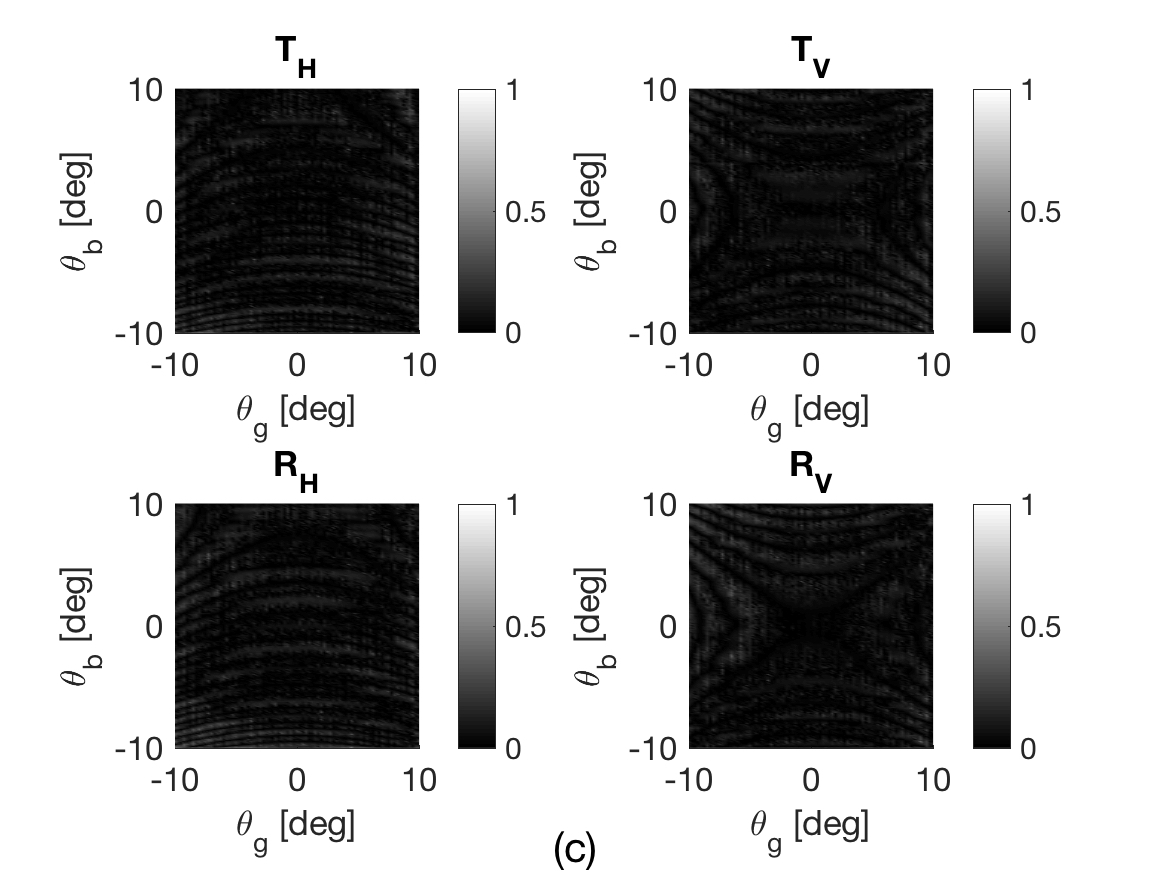} 
\caption{(a) Experimental transmission $T$ and reflection $R$ coefficients
for the horizontal $H$ and vertical $V$ polarizations in the output
ports of the variable partially-polarizing beam splitter shown in
Fig. \ref{ExpSetup}. The axes, $\theta_{b}$ and $\theta_{g}$, are
the tilts of the birefrigent crystal and glass plate, respectively.
(b) Theoretically expected $T$ and $R$ coefficients
for the $H$ and $V$ polarizations of the variable partially-polarizing
beam splitter in Fig. \ref{ExpSetup} according to the theoretical
model in Sec. \ref{Theory} and the fitting procedure explained in Sec. \ref{ExpResults}. (c) Absolute difference between experiment and theory. The maximum absolute difference for each coefficient is 0.36 ($T_H$), 0.39 ($T_V$), 0.43 ($R_H$), and 0.41 ($R_V$). The dashed and dotted lines in (a) and (b) correspond to the bands used to illustrate a complete and independent control
of the variable partially-polarizing beam splitter coefficients in Fig. \ref{2DPlots}.}
\label{data1} 
\end{figure}
\par\end{center}
\twocolumngrid
\ \\

The first observation in Fig. \ref{data1}(a) is that the VPPBS coefficients
exhibit several maxima and minima in the range of $\theta_{b}$ and
$\theta_{g}$ values studied here. At these points there is either
constructive or destructive interference due to the relative phases
introduced by the birefringent crystal and glass plate. As expected,
the $T_{H}$ and $R_{H}$ coefficients are complementary to one another,
i.e. when $T_{H}$ is maximum $R_{H}$ is minimum and vice versa.
This complementarity happens as well for $T_{V}$ and $R_{V}$. Furthermore,
the coefficients in Fig. \ref{data1}(a) are symmetric with respect
to the axis $\theta_{g}=0$, which comes from the fact that the relative
phase introduced by the glass plate is the same for both positive
and negative $\theta_{g}$ angles. The same happens for $T_{V}$ and
$R_{V}$ with respect to the axis $\theta_{b}=0$ since $n_{b}^{V}$
is a constant. In contrast, $T_{H}$ and $R_{H}$ are asymmetric with
respect to the axis $\theta_{b}=0$ because the angle $\theta_{c}$
of the optic axis $\vec{c}$ is not zero when the face of the birefringent
crystal is perpendicular to the beam. This creates a difference between
a positive or negative tilt of the crystal that only affects the horizontal
polarization, in accordance with Eq. (\ref{nbH}). 

In order to compare the experimental results in Fig. \ref{data1}(a)
with the theoretical model in Sec. \ref{Theory}, Eqs. (\ref{TH})-(\ref{nbH}), a least-squares fitting of the experimental data is performed in the following way. First, the experimental visibilities for the four coefficients in Fig. \ref{data1}(a) are taken into account in the theoretical model by using a modified version of Eq. (\ref{TH}),
\begin{equation}
T_\epsilon=\frac{1+\mathcal{V}_\epsilon \cos \phi_\epsilon}{2},
\label{Visibility}
\end{equation}

\noindent where $\mathcal{V}_\epsilon$ is the experimental visibility for the coefficient $T_\epsilon$ ($\epsilon = H, V$.) The reflection coefficients are still given by $R_\epsilon=1-T_\epsilon$. Second, the thicknesses $d_b$ and $d_g$ for the phase retarders are adjusted iteratively using only $T_V$ in Fig. \ref{data1}(a) since the index of refraction for the $V$ polarization is a constant and therefore independent of $\theta_c$, the third fitted parameter described below ($R_V$ could have been used as well). The adjusted values for $d_b$ and $d_g$ are 0.2724 mm and 0.1477 mm, respectively. Third, using these adjusted thicknesses, the experimental coefficient $T_H$ is iteratively fitted using $\theta_c$ and a tilt angle offset $\theta_0$ for the birefringent crystal (such that $\theta_b\to\theta_b+\theta_0$) as fitting parameters. The results for $\theta_c$ and $\theta_0$ are respectively $32.14^\circ$ and $0.061^\circ$. Finally, the four adjusted parameters are introduced in the theoretical model and the resulting coefficients are depicted in Fig. \ref{data1}(b). The values for the refractive indices are $n_{g}=1.5151$ (BK7 refractive index \cite{Schott}), and $n_{o}=1.6672$ and $n_{e}=1.5496$ (from the Sellmeier's equations \cite{Kato86} for BBO at $\lambda=632.8$ nm).

In order to easily compare the experimental results to the theory in Fig. \ref{data1}, a plot showing the absolute difference between the two is presented in Fig. \ref{data1}(c). Qualitatively, the theoretical model reproduces the experimental VPPBS performance. However, even after all the careful fitting described above, the maximum absolute difference, 0.43, is remarkably large. The refractive indices are the only parameters that were not fitted, which suggests that they may a contributor to this discrepancy. In any case, we conclude that the theoretical model is insufficiently accurate to predict the required tilt angles for a desired transmission coefficient. Instead, the experimental characterization in Fig. \ref{data1}(a) must be used.

\begin{figure}[htbp]
\centering \includegraphics[width=\linewidth]{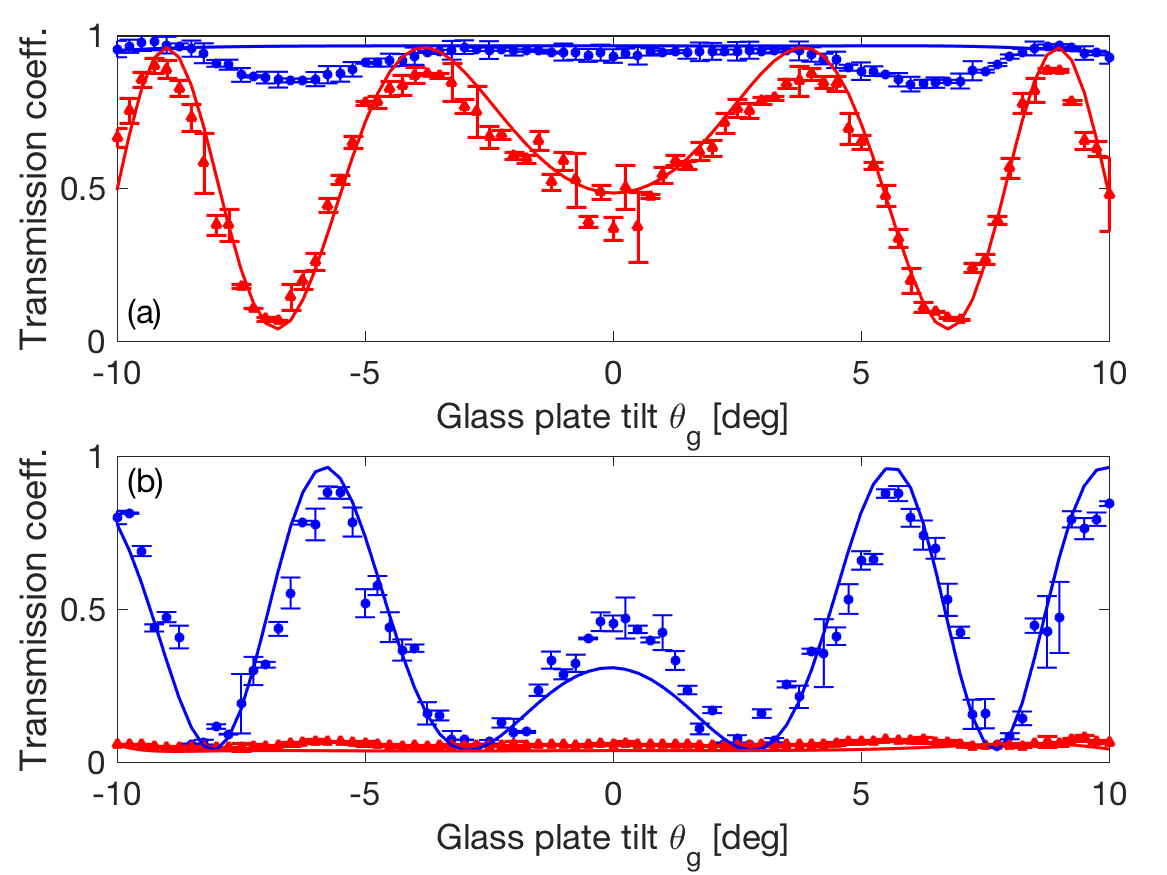} \caption{Transmission coefficients for the horizontal (blue circles) and vertical
(red triangles) polarizations in two particular cases: (a) when $T_{H}$
is maximum and $T_{V}$ varies, and (b) when $T_{V}$ is minimum and
$T_{H}$ varies. These two plots respectively correspond to the dashed
and dotted lines in Fig. \ref{data1}(a). The solid curves correspond to the theoretical transmission coefficients along the same two lines. In order to be visible, the error bars correspond
to three times the standard deviation over the 100 measurements taken for
each data point. The noise observed is mainly due to laser instability
arising from back reflections.}
\label{2DPlots} 
\end{figure}

The VPPBS working principle can be illustrated using Fig. \ref{data1}(a)
by finding the tilt angles at which one of the transmission coefficients
is kept constant while the other varies. As example, it is shown here
the case when $T_{H}$ is kept constant while $T_{V}$ varies. This
can be accomplished by selecting any band in the $T_{H}$ plot for
which this coefficient is constant, e.g. the one marked with a dashed
(blue) line in Fig. \ref{data1}(a). For the same tilt angles that
describe such a line in the $T_{H}$ contour plot, a set of values
between 0 and 1 is found for the $T_{V}$ coefficient, as indicated by a dashed (red) line in Fig. \ref{data1}(a). The latter case is illustrated
in Fig. \ref{2DPlots}(a), where the $T_{H}$ and $T_{V}$ coefficients
are shown as a function of $\theta_{g}$ that parametrizes the dashed
line in Fig. \ref{data1}(a). In Fig. \ref{2DPlots}(b) a second case
is considered, $T_{V}$ set to its minimum value while $T_{H}$ is
varied. One of the bands in the plot for $T_{V}$ in Fig. \ref{data1}(a)
that fulfills this condition is highlighted with a dotted (red)
line. In this case, the $T_{H}$ coefficient achieves values between
0 and 1 indicated by a dotted (blue) line in Fig. \ref{data1}(a). Therefore, by selecting an appropriate value for $\theta_{g}$ that parametrizes the dotted line, it is possible to get an arbitrary
value for $T_{H}$. 

The two cases summarized in Fig. \ref{2DPlots} illustrate the fact
that the $T_{H}$ and $T_{V}$ coefficients, and therefore their reflection
counterparts, can be controlled at will by choosing two tilt angles.
In other words, the relative phases introduced by the birefringent
crystal and glass plate allow a complete and independent control of
the $T$ and $R$ coefficients for both polarizations, which is the
defining feature of a VPPBS.

The performance of the VPPBS presented here can be quantified in terms
of the interferometer visibility for each polarization. On one hand,
the overall visibility for the $H$ polarization ($T_{H}$ coefficient
in Fig. \ref{data1}(a)) was 93\%, while for the $V$ polarization
($T_{V}$ coefficient in Fig. \ref{data1}(a)) was 92\%. In the particular
cases shown in Fig. \ref{2DPlots}, the $T_{V}$ (panel (a)) and $T_{H}$
(panel (b)) coefficients display visibilities of 89\% and 86\%, respectively.
We have considered and ruled out a number of possible sources for the imperfect visibility, including imbalanced NPBS splitting ratios, fluctuations in the signal, and polarization. This leaves the most likely source to be alignment of the interferometer. In any case, the main impact of imperfect visibility will be to limit the achievable range of the transmission coefficient, as can be seen in Eq. (\ref{Visibility}). Despite this fact, the Sagnac interferometer remained stable over 12 hours without the need for active feedback or constant readjustment, which suggests that the VPPBS could successfully be used as an element in a larger
experimental setup.

Lastly, we discuss the theoretical lines in Fig. \ref{2DPlots}.  Qualitatively, the strongly varying transmission coefficients in panels (a) (red curve) and (b) (blue curve) follow the behaviour of the experimental points. However, the transmission coefficients that are meant to be constant do vary unlike their corresponding theoretical curves. Moreover, for all four curves, the discrepancy between theory and experiment is greater than the experimental uncertainty for most data points. Given the low accuracy of the theoretical model and its relative complexity, this discrepancy again confirms that it is better to use the experimental characterization of the VPPBS presented in Fig. \ref{data1}(a) to determine the correct tilt media angles for a target transmission coefficient.

\section{\label{Conclusions}Conclusions}

A variable partially-polarizing beam splitter is presented based on
a displaced Sagnac interferometer. The transmission and reflection
coefficients for the horizontal and vertical polarizations are controlled
via the tilts of a birefringent crystal and a glass plate, which introduce
a relative phase to each polarization. The overall effect of these
two phase retarders is a complete and independent manipulation of
the VPPBS splitting ratios for the two polarizations. Since this design
includes optical elements that can be found in any optics laboratory, its
implementation is straightforward and inexpensive.

This work was supported by the Canada Research Chairs (CRC) Program,
the Natural Sciences and Engineering Research Council (NSERC), and
the Cananda Excellence Research Chairs (CERC) Program. JF acknowledges
support from COLCIENCIAS.

\bibliography{References}

\begin{thebibliography}{16}%
\makeatletter
\providecommand \@ifxundefined [1]{%
 \@ifx{#1\undefined}
}%
\providecommand \@ifnum [1]{%
 \ifnum #1\expandafter \@firstoftwo
 \else \expandafter \@secondoftwo
 \fi
}%
\providecommand \@ifx [1]{%
 \ifx #1\expandafter \@firstoftwo
 \else \expandafter \@secondoftwo
 \fi
}%
\providecommand \natexlab [1]{#1}%
\providecommand \enquote  [1]{``#1''}%
\providecommand \bibnamefont  [1]{#1}%
\providecommand \bibfnamefont [1]{#1}%
\providecommand \citenamefont [1]{#1}%
\providecommand \href@noop [0]{\@secondoftwo}%
\providecommand \href [0]{\begingroup \@sanitize@url \@href}%
\providecommand \@href[1]{\@@startlink{#1}\@@href}%
\providecommand \@@href[1]{\endgroup#1\@@endlink}%
\providecommand \@sanitize@url [0]{\catcode `\\12\catcode `\$12\catcode
  `\&12\catcode `\#12\catcode `\^12\catcode `\_12\catcode `\%12\relax}%
\providecommand \@@startlink[1]{}%
\providecommand \@@endlink[0]{}%
\providecommand \url  [0]{\begingroup\@sanitize@url \@url }%
\providecommand \@url [1]{\endgroup\@href {#1}{\urlprefix }}%
\providecommand \urlprefix  [0]{URL }%
\providecommand \Eprint [0]{\href }%
\providecommand \doibase [0]{http://dx.doi.org/}%
\providecommand \selectlanguage [0]{\@gobble}%
\providecommand \bibinfo  [0]{\@secondoftwo}%
\providecommand \bibfield  [0]{\@secondoftwo}%
\providecommand \translation [1]{[#1]}%
\providecommand \BibitemOpen [0]{}%
\providecommand \bibitemStop [0]{}%
\providecommand \bibitemNoStop [0]{.\EOS\space}%
\providecommand \EOS [0]{\spacefactor3000\relax}%
\providecommand \BibitemShut  [1]{\csname bibitem#1\endcsname}%
\let\auto@bib@innerbib\@empty
\bibitem [{\citenamefont {Kiesel}\ \emph {et~al.}(2005)\citenamefont {Kiesel},
  \citenamefont {Schmid}, \citenamefont {Weber}, \citenamefont {Ursin},\ and\
  \citenamefont {Weinfurter}}]{Kiesel05}%
  \BibitemOpen
  \bibfield  {author} {\bibinfo {author} {\bibfnamefont {N.}~\bibnamefont
  {Kiesel}}, \bibinfo {author} {\bibfnamefont {C.}~\bibnamefont {Schmid}},
  \bibinfo {author} {\bibfnamefont {U.}~\bibnamefont {Weber}}, \bibinfo
  {author} {\bibfnamefont {R.}~\bibnamefont {Ursin}}, \ and\ \bibinfo {author}
  {\bibfnamefont {H.}~\bibnamefont {Weinfurter}},\ }\bibfield  {title}
  {\enquote {\bibinfo {title} {Linear optics controlled-phase gate made
  simple},}\ }\href {\doibase 10.1103/PhysRevLett.95.210505} {\bibfield
  {journal} {\bibinfo  {journal} {Phys. Rev. Lett.}\ }\textbf {\bibinfo
  {volume} {95}},\ \bibinfo {pages} {210505} (\bibinfo {year}
  {2005})}\BibitemShut {NoStop}%
\bibitem [{\citenamefont {Okamoto}\ \emph {et~al.}(2005)\citenamefont
  {Okamoto}, \citenamefont {Hofmann}, \citenamefont {Takeuchi},\ and\
  \citenamefont {Sasaki}}]{Okamoto05}%
  \BibitemOpen
  \bibfield  {author} {\bibinfo {author} {\bibfnamefont {R.}~\bibnamefont
  {Okamoto}}, \bibinfo {author} {\bibfnamefont {H.~F.}\ \bibnamefont
  {Hofmann}}, \bibinfo {author} {\bibfnamefont {S.}~\bibnamefont {Takeuchi}}, \
  and\ \bibinfo {author} {\bibfnamefont {K.}~\bibnamefont {Sasaki}},\
  }\bibfield  {title} {\enquote {\bibinfo {title} {Demonstration of an optical
  quantum controlled-not gate without path interference},}\ }\href {\doibase
  10.1103/PhysRevLett.95.210506} {\bibfield  {journal} {\bibinfo  {journal}
  {Phys. Rev. Lett.}\ }\textbf {\bibinfo {volume} {95}},\ \bibinfo {pages}
  {210506} (\bibinfo {year} {2005})}\BibitemShut {NoStop}%
\bibitem [{\citenamefont {Langford}\ \emph {et~al.}(2005)\citenamefont
  {Langford}, \citenamefont {Weinhold}, \citenamefont {Prevedel}, \citenamefont
  {Resch}, \citenamefont {Gilchrist}, \citenamefont {O'Brien}, \citenamefont
  {Pryde},\ and\ \citenamefont {White}}]{Langford05}%
  \BibitemOpen
  \bibfield  {author} {\bibinfo {author} {\bibfnamefont {N.~K.}\ \bibnamefont
  {Langford}}, \bibinfo {author} {\bibfnamefont {T.~J.}\ \bibnamefont
  {Weinhold}}, \bibinfo {author} {\bibfnamefont {R.}~\bibnamefont {Prevedel}},
  \bibinfo {author} {\bibfnamefont {K.~J.}\ \bibnamefont {Resch}}, \bibinfo
  {author} {\bibfnamefont {A.}~\bibnamefont {Gilchrist}}, \bibinfo {author}
  {\bibfnamefont {J.~L.}\ \bibnamefont {O'Brien}}, \bibinfo {author}
  {\bibfnamefont {G.~J.}\ \bibnamefont {Pryde}}, \ and\ \bibinfo {author}
  {\bibfnamefont {A.~G.}\ \bibnamefont {White}},\ }\bibfield  {title} {\enquote
  {\bibinfo {title} {Demonstration of a simple entangling optical gate and its
  use in {B}ell-state analysis},}\ }\href {\doibase
  10.1103/PhysRevLett.95.210504} {\bibfield  {journal} {\bibinfo  {journal}
  {Phys. Rev. Lett.}\ }\textbf {\bibinfo {volume} {95}},\ \bibinfo {pages}
  {210504} (\bibinfo {year} {2005})}\BibitemShut {NoStop}%
\bibitem [{\citenamefont {Ling}\ \emph {et~al.}(2006)\citenamefont {Ling},
  \citenamefont {Soh}, \citenamefont {Lamas-Linares},\ and\ \citenamefont
  {Kurtsiefer}}]{Ling06}%
  \BibitemOpen
  \bibfield  {author} {\bibinfo {author} {\bibfnamefont {A.}~\bibnamefont
  {Ling}}, \bibinfo {author} {\bibfnamefont {K.~P.}\ \bibnamefont {Soh}},
  \bibinfo {author} {\bibfnamefont {A.}~\bibnamefont {Lamas-Linares}}, \ and\
  \bibinfo {author} {\bibfnamefont {C.}~\bibnamefont {Kurtsiefer}},\ }\bibfield
   {title} {\enquote {\bibinfo {title} {Experimental polarization state
  tomography using optimal polarimeters},}\ }\href {\doibase
  10.1103/PhysRevA.74.022309} {\bibfield  {journal} {\bibinfo  {journal} {Phys.
  Rev. A}\ }\textbf {\bibinfo {volume} {74}},\ \bibinfo {pages} {022309}
  (\bibinfo {year} {2006})}\BibitemShut {NoStop}%
\bibitem [{\citenamefont {Medendorp}\ \emph {et~al.}(2011)\citenamefont
  {Medendorp}, \citenamefont {Torres-Ruiz}, \citenamefont {Shalm},
  \citenamefont {Tabia}, \citenamefont {Fuchs},\ and\ \citenamefont
  {Steinberg}}]{Medendorp11}%
  \BibitemOpen
  \bibfield  {author} {\bibinfo {author} {\bibfnamefont {Z.~E.~D.}\
  \bibnamefont {Medendorp}}, \bibinfo {author} {\bibfnamefont {F.~A.}\
  \bibnamefont {Torres-Ruiz}}, \bibinfo {author} {\bibfnamefont {L.~K.}\
  \bibnamefont {Shalm}}, \bibinfo {author} {\bibfnamefont {G.~N.~M.}\
  \bibnamefont {Tabia}}, \bibinfo {author} {\bibfnamefont {C.~A.}\ \bibnamefont
  {Fuchs}}, \ and\ \bibinfo {author} {\bibfnamefont {A.~M.}\ \bibnamefont
  {Steinberg}},\ }\bibfield  {title} {\enquote {\bibinfo {title} {Experimental
  characterization of qutrits using symmetric informationally complete positive
  operator-valued measurements},}\ }\href {\doibase 10.1103/PhysRevA.83.051801}
  {\bibfield  {journal} {\bibinfo  {journal} {Phys. Rev. A}\ }\textbf {\bibinfo
  {volume} {83}},\ \bibinfo {pages} {051801} (\bibinfo {year}
  {2011})}\BibitemShut {NoStop}%
\bibitem [{\citenamefont {Kaiser}\ \emph {et~al.}(2012)\citenamefont {Kaiser},
  \citenamefont {Coudreau}, \citenamefont {Milman}, \citenamefont {Ostrowsky},\
  and\ \citenamefont {Tanzilli}}]{Kaiser12}%
  \BibitemOpen
  \bibfield  {author} {\bibinfo {author} {\bibfnamefont {F.}~\bibnamefont
  {Kaiser}}, \bibinfo {author} {\bibfnamefont {T.}~\bibnamefont {Coudreau}},
  \bibinfo {author} {\bibfnamefont {P.}~\bibnamefont {Milman}}, \bibinfo
  {author} {\bibfnamefont {D.~B.}\ \bibnamefont {Ostrowsky}}, \ and\ \bibinfo
  {author} {\bibfnamefont {S.}~\bibnamefont {Tanzilli}},\ }\bibfield  {title}
  {\enquote {\bibinfo {title} {Entanglement-enabled delayed-choice
  experiment},}\ }\href {\doibase 10.1126/science.1226755} {\bibfield
  {journal} {\bibinfo  {journal} {Science}\ }\textbf {\bibinfo {volume}
  {338}},\ \bibinfo {pages} {637} (\bibinfo {year} {2012})}\BibitemShut
  {NoStop}%
\bibitem [{\citenamefont {Boyd}(2008)}]{Boyd}%
  \BibitemOpen
  \bibfield  {author} {\bibinfo {author} {\bibfnamefont {R.~W.}\ \bibnamefont
  {Boyd}},\ }\href@noop {} {\emph {\bibinfo {title} {Nonlinear Optics, Third
  Edition}}},\ \bibinfo {edition} {3rd}\ ed.\ (\bibinfo  {publisher} {Academic
  Press},\ \bibinfo {year} {2008})\BibitemShut {NoStop}%
\bibitem [{\citenamefont {Nagata}\ \emph {et~al.}(2007)\citenamefont {Nagata},
  \citenamefont {Okamoto}, \citenamefont {O'Brien}, \citenamefont {Sasaki},\
  and\ \citenamefont {Takeuchi}}]{Nagata07}%
  \BibitemOpen
  \bibfield  {author} {\bibinfo {author} {\bibfnamefont {T.}~\bibnamefont
  {Nagata}}, \bibinfo {author} {\bibfnamefont {R.}~\bibnamefont {Okamoto}},
  \bibinfo {author} {\bibfnamefont {J.~L.}\ \bibnamefont {O'Brien}}, \bibinfo
  {author} {\bibfnamefont {K.}~\bibnamefont {Sasaki}}, \ and\ \bibinfo {author}
  {\bibfnamefont {S.}~\bibnamefont {Takeuchi}},\ }\bibfield  {title} {\enquote
  {\bibinfo {title} {Beating the standard quantum limit with four-entangled
  photons},}\ }\href {\doibase 10.1126/science.1138007} {\bibfield  {journal}
  {\bibinfo  {journal} {Science}\ }\textbf {\bibinfo {volume} {316}},\ \bibinfo
  {pages} {726} (\bibinfo {year} {2007})}\BibitemShut {NoStop}%
\bibitem [{\citenamefont {Mi{\v c}uda}\ \emph {et~al.}(2014)\citenamefont
  {Mi{\v c}uda}, \citenamefont {Dol{\'a}kov{\'a}}, \citenamefont {Straka},
  \citenamefont {Mikov{\'a}}, \citenamefont {Du{\v s}ek}, \citenamefont
  {Fiur{\'a}{\v s}ek},\ and\ \citenamefont {Je{\v z}ek}}]{Micuda14}%
  \BibitemOpen
  \bibfield  {author} {\bibinfo {author} {\bibfnamefont {M.}~\bibnamefont
  {Mi{\v c}uda}}, \bibinfo {author} {\bibfnamefont {E.}~\bibnamefont
  {Dol{\'a}kov{\'a}}}, \bibinfo {author} {\bibfnamefont {I.}~\bibnamefont
  {Straka}}, \bibinfo {author} {\bibfnamefont {M.}~\bibnamefont {Mikov{\'a}}},
  \bibinfo {author} {\bibfnamefont {M.}~\bibnamefont {Du{\v s}ek}}, \bibinfo
  {author} {\bibfnamefont {J.}~\bibnamefont {Fiur{\'a}{\v s}ek}}, \ and\
  \bibinfo {author} {\bibfnamefont {M.}~\bibnamefont {Je{\v z}ek}},\ }\bibfield
   {title} {\enquote {\bibinfo {title} {Highly stable polarization independent
  {M}ach-{Z}ehnder interferometer},}\ }\href {\doibase
  http://dx.doi.org/10.1063/1.4891702} {\bibfield  {journal} {\bibinfo
  {journal} {Review of Scientific Instruments}\ }\textbf {\bibinfo {volume}
  {85}},\ \bibinfo {eid} {083103} (\bibinfo {year} {2014})}\BibitemShut
  {NoStop}%
\bibitem [{\citenamefont {Ashby}, \citenamefont {Schwarz},\ and\ \citenamefont
  {Schlosshauer}(2016)}]{Ashby16}%
  \BibitemOpen
  \bibfield  {author} {\bibinfo {author} {\bibfnamefont {J.~M.}\ \bibnamefont
  {Ashby}}, \bibinfo {author} {\bibfnamefont {P.~D.}\ \bibnamefont {Schwarz}},
  \ and\ \bibinfo {author} {\bibfnamefont {M.}~\bibnamefont {Schlosshauer}},\
  }\bibfield  {title} {\enquote {\bibinfo {title} {Observation of the quantum
  paradox of separation of a single photon from one of its properties},}\
  }\href {\doibase 10.1103/PhysRevA.94.012102} {\bibfield  {journal} {\bibinfo
  {journal} {Physical Review A}\ }\textbf {\bibinfo {volume} {94}},\ \bibinfo
  {pages} {012102} (\bibinfo {year} {2016})}\BibitemShut {NoStop}%
\bibitem [{\citenamefont {Schott}(2015)}]{Schott}%
  \BibitemOpen
  \bibfield  {author} {\bibinfo {author} {\bibnamefont {Schott}},\ }\href@noop
  {} {\enquote {\bibinfo {title} {Optical glass data sheets},}\ } (\bibinfo
  {year} {2015})\BibitemShut {NoStop}%
\bibitem [{\citenamefont {Kato}(1986)}]{Kato86}%
  \BibitemOpen
  \bibfield  {author} {\bibinfo {author} {\bibfnamefont {K.}~\bibnamefont
  {Kato}},\ }\bibfield  {title} {\enquote {\bibinfo {title} {Second-harmonic
  generation to 2048 {A} in beta-{B}a{B}2{O}4},}\ }\href {\doibase
  10.1109/JQE.1986.1073097} {\bibfield  {journal} {\bibinfo  {journal} {IEEE
  Journal of Quantum Electronics}\ }\textbf {\bibinfo {volume} {QE-22}},\
  \bibinfo {pages} {1013} (\bibinfo {year} {1986})}\BibitemShut {NoStop}%
\bibitem [{\citenamefont {Jeong}, \citenamefont {Lee},\ and\ \citenamefont
  {Kim}(2013)}]{Jeong13}%
  \BibitemOpen
  \bibfield  {author} {\bibinfo {author} {\bibfnamefont {Y.-C.}\ \bibnamefont
  {Jeong}}, \bibinfo {author} {\bibfnamefont {J.-C.}\ \bibnamefont {Lee}}, \
  and\ \bibinfo {author} {\bibfnamefont {Y.-H.}\ \bibnamefont {Kim}},\
  }\bibfield  {title} {\enquote {\bibinfo {title} {Experimental implementation
  of a fully controllable depolarizing quantum operation},}\ }\href {\doibase
  10.1103/PhysRevA.87.014301} {\bibfield  {journal} {\bibinfo  {journal}
  {Physical Review A}\ }\textbf {\bibinfo {volume} {87}},\ \bibinfo {pages}
  {014301} (\bibinfo {year} {2013})}\BibitemShut {NoStop}%
\bibitem [{\citenamefont {Cuevas}\ \emph {et~al.}(2017)\citenamefont {Cuevas},
  \citenamefont {Proietti}, \citenamefont {Ciampini}, \citenamefont {Duranti},
  \citenamefont {Mataloni}, \citenamefont {Sacchi},\ and\ \citenamefont
  {Macchiavello}}]{Cuevas17}%
  \BibitemOpen
  \bibfield  {author} {\bibinfo {author} {\bibfnamefont {A.}~\bibnamefont
  {Cuevas}}, \bibinfo {author} {\bibfnamefont {M.}~\bibnamefont {Proietti}},
  \bibinfo {author} {\bibfnamefont {M.~A.}\ \bibnamefont {Ciampini}}, \bibinfo
  {author} {\bibfnamefont {S.}~\bibnamefont {Duranti}}, \bibinfo {author}
  {\bibfnamefont {P.}~\bibnamefont {Mataloni}}, \bibinfo {author}
  {\bibfnamefont {M.~F.}\ \bibnamefont {Sacchi}}, \ and\ \bibinfo {author}
  {\bibfnamefont {C.}~\bibnamefont {Macchiavello}},\ }\bibfield  {title}
  {\enquote {\bibinfo {title} {Experimental detection of quantum channel
  capacities},}\ }\href {\doibase 10.1103/PhysRevLett.119.100502} {\bibfield
  {journal} {\bibinfo  {journal} {Physical Review Letters}\ }\textbf {\bibinfo
  {volume} {119}},\ \bibinfo {pages} {100502} (\bibinfo {year}
  {2017})}\BibitemShut {NoStop}%
\bibitem [{\citenamefont {O'Brien}\ \emph {et~al.}(2003)\citenamefont
  {O'Brien}, \citenamefont {Pryde}, \citenamefont {White}, \citenamefont
  {Ralph},\ and\ \citenamefont {Branning}}]{OBrien03}%
  \BibitemOpen
  \bibfield  {author} {\bibinfo {author} {\bibfnamefont {J.~L.}\ \bibnamefont
  {O'Brien}}, \bibinfo {author} {\bibfnamefont {G.~J.}\ \bibnamefont {Pryde}},
  \bibinfo {author} {\bibfnamefont {A.~G.}\ \bibnamefont {White}}, \bibinfo
  {author} {\bibfnamefont {T.~C.}\ \bibnamefont {Ralph}}, \ and\ \bibinfo
  {author} {\bibfnamefont {D.}~\bibnamefont {Branning}},\ }\bibfield  {title}
  {\enquote {\bibinfo {title} {Demonstration of an all-optical quantum
  controlled-not gate},}\ }\href {http://dx.doi.org/10.1038/nature02054}
  {\bibfield  {journal} {\bibinfo  {journal} {Nature}\ }\textbf {\bibinfo
  {volume} {426}},\ \bibinfo {pages} {264} (\bibinfo {year}
  {2003})}\BibitemShut {NoStop}%
\bibitem [{\citenamefont {Evans}\ \emph {et~al.}(2010)\citenamefont {Evans},
  \citenamefont {Bennink}, \citenamefont {Grice}, \citenamefont {Humble},\ and\
  \citenamefont {Schaake}}]{Evans10}%
  \BibitemOpen
  \bibfield  {author} {\bibinfo {author} {\bibfnamefont {P.~G.}\ \bibnamefont
  {Evans}}, \bibinfo {author} {\bibfnamefont {R.~S.}\ \bibnamefont {Bennink}},
  \bibinfo {author} {\bibfnamefont {W.~P.}\ \bibnamefont {Grice}}, \bibinfo
  {author} {\bibfnamefont {T.~S.}\ \bibnamefont {Humble}}, \ and\ \bibinfo
  {author} {\bibfnamefont {J.}~\bibnamefont {Schaake}},\ }\bibfield  {title}
  {\enquote {\bibinfo {title} {Bright source of spectrally uncorrelated
  polarization-entangled photons with nearly single-mode emission},}\ }\href
  {\doibase 10.1103/PhysRevLett.105.253601} {\bibfield  {journal} {\bibinfo
  {journal} {Phys. Rev. Lett.}\ }\textbf {\bibinfo {volume} {105}},\ \bibinfo
  {pages} {253601} (\bibinfo {year} {2010})}\BibitemShut {NoStop}%
\end{thebibliography}%

\appendix
\section{Alternative VPPBS configurations} \label{OtherVPPBSs} 

\begin{figure}[htbp]
\centering \includegraphics[width=\linewidth]{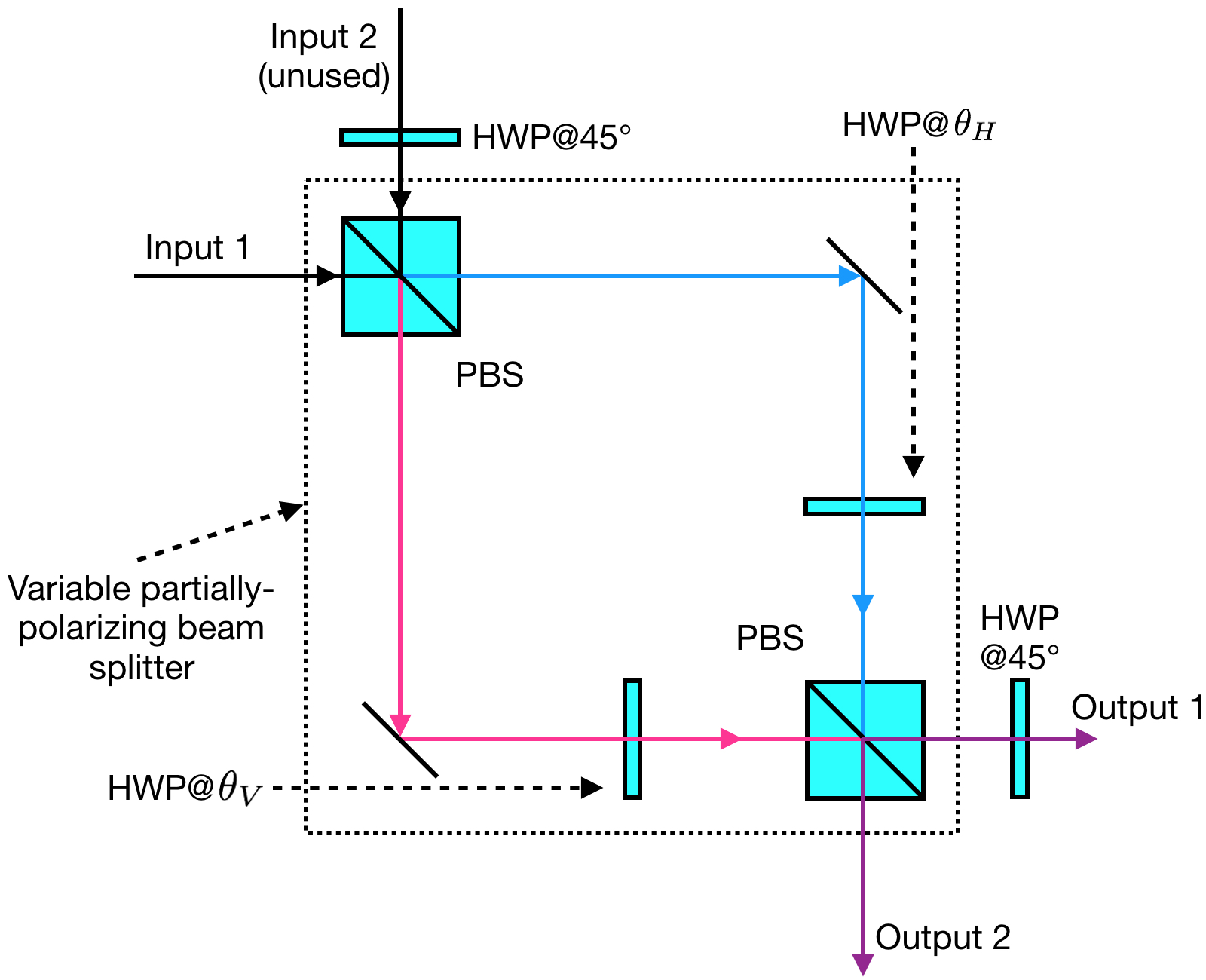} \caption{Variable partially-polarizing beam splitter based on a Mach-Zehnder
interferometer with polarizing beam splitters (PBSs) instead of non-polarizing
beam splitters as in Fig. \ref{MZ}, and half-wave plates (HWPs) as
phase retarders. Although Input 2 is unused in the current description
of the VPPBS, a HWP at 45$^{\circ}$ must be placed at this input
to convert the horizontal polarization at Input 2 to vertical so that it travels into the same arm as the horizontal polarization at Input 1. Consequently, the two inputs can then be mixed by the HWP in that arm. A similar argument holds for the vertical polarizations at Input 1 and 2. This HWP at Input 2 is necessary when both inputs of the VPPBS are required, as in two-photon quantum logic gate operations\cite{Kiesel05,Okamoto05,Langford05,OBrien03}.}
\label{MZPBS} 
\end{figure}

A VPPBS can also be created using a variation
of the Mach-Zehnder interferometer in Fig. \ref{MZ}. As shown
in Fig. \ref{MZPBS}, by replacing the NPBSs by PBSs and implementing
the phase retarders via half-wave plates (HWPs), similar expressions
for $\mathbf{E}_{\text{out,1}}(t)$ and $\mathbf{E}_{\text{out,1}}(t)$
in Eqs. (\ref{Eout1}) and (\ref{Eout2}) are obtained. Indeed, the interferometer
in Fig. \ref{MZPBS} has its own displaced Sagnac-interferometer version 
presented in Refs. \cite{Jeong13,Cuevas17}, except for the HWP at 45$^{\circ}$
in one of its outputs. The main reason to implement experimentally the Sagnac interferometer in Fig. \ref{ExpSetup} instead of the one in Refs. \cite{Jeong13,Cuevas17} is that the separation between the counter-propagating paths does not provide enough room to place the HWP rotating mounts at our disposal without blocking one of the beams. 

To see explicitly how the Mach-Zehnder interferometer in Fig. \ref{MZPBS}
works as a VPPBS, consider light entering at Input 1 with the electric
field $\mathbf{E}_{\text{in}}(t)$ in Eq. \ref{Ein}. After the first
PBS the upper and lower fields read 
\begin{gather}
\mathbf{E}_{\text{upper}}(t)=\left[\begin{array}{c}
E_{\text{in}}^{H}(t)\\
\\
0
\end{array}\right],\\
\nonumber \\
\mathbf{E}_{\text{lower}}(t)=e^{i\frac{\pi}{2}}\left[\begin{array}{c}
0\\
\\
E_{\text{in}}^{V}(t)
\end{array}\right].
\end{gather}

After the HWPs with fast axes oriented at $\theta_{H}$ and $\theta_{V}$
in the upper and lower paths, respectively, the electric fields become
\begin{gather}
\tilde{\mathbf{E}}_{\text{upper}}(t)=E_{\text{in}}^{H}(t)\left[\begin{array}{c}
\cos(2\theta_{H})\\
\\
\sin(2\theta_{H})
\end{array}\right],\\
\nonumber \\
\tilde{\mathbf{E}}_{\text{lower}}(t)=e^{i\frac{\pi}{2}}E_{\text{in}}^{V}(t)\left[\begin{array}{c}
\sin(2\theta_{V})\\
\\
-\cos(2\theta_{V})
\end{array}\right].
\end{gather}

Finally, at Outputs 1 and 2, including the HWP at 45$^{\circ}$ in
the first one, the electric fields are 
\begin{gather}
\mathbf{E}_{\text{out,1}}(t)=\left[\begin{array}{c}
E_{\text{in}}^{H}(t)\sin(2\theta_{H})\\
\\
E_{\text{in}}^{V}(t)\sin(2\theta_{V})
\end{array}\right],\\
\nonumber \\
\mathbf{E}_{\text{out,2}}(t)=\left[\begin{array}{c}
E_{\text{in}}^{H}(t)\cos(2\theta_{H})\\
\\
E_{\text{in}}^{V}(t)\cos(2\theta_{V})
\end{array}\right].
\end{gather}

These expressions are physically identical to Eqs. (\ref{Eout1})
and (\ref{Eout2}), and therefore allow the definition of the VPPBS
transmission and reflection coefficients as in Eq. (\ref{TH}). 

\begin{figure}[htbp]
\centering \includegraphics[width=\linewidth]{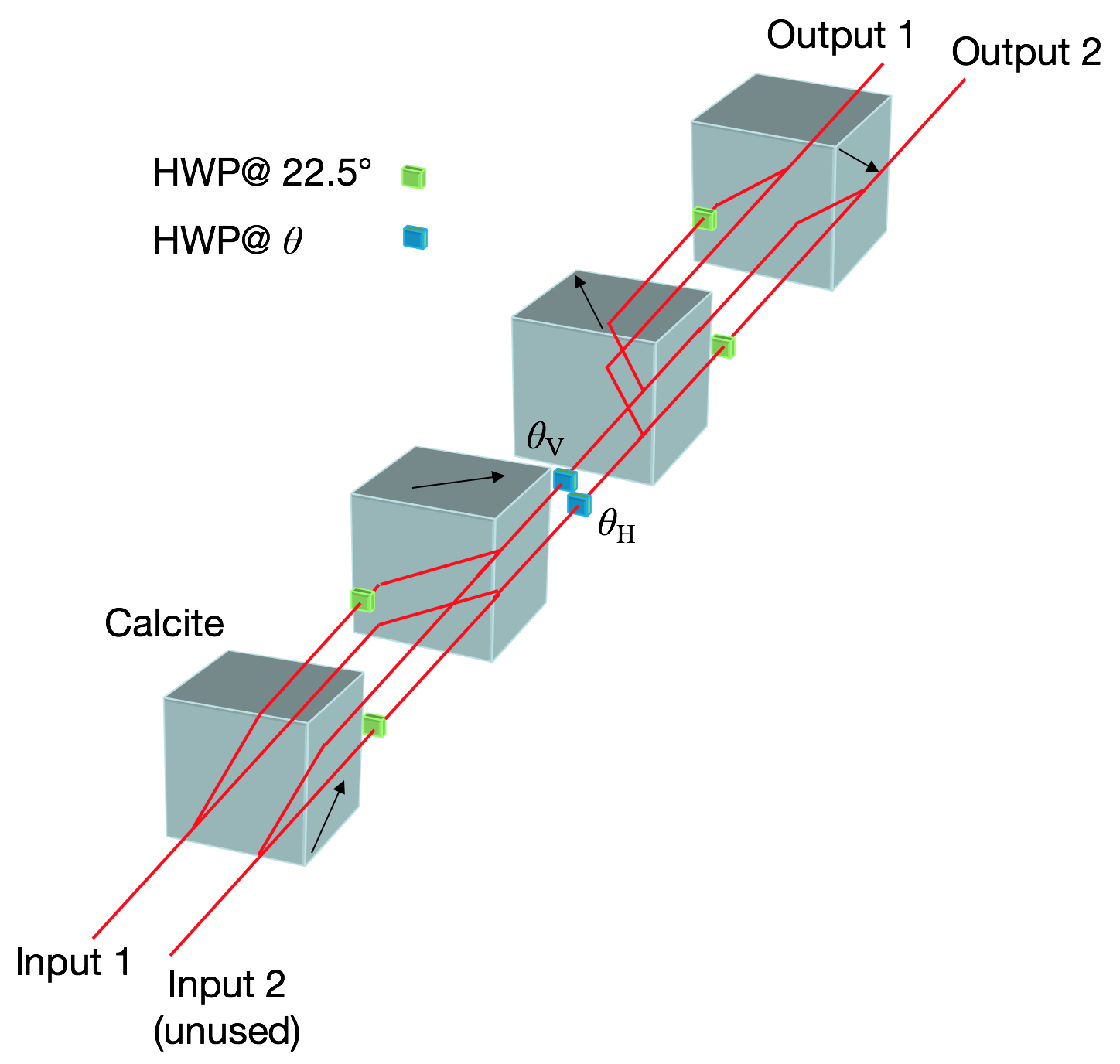}
\caption{Alternative VPPBS configuration using PBSs made of birefringent crystals in a linear configuration. The resulting interferometer exhibits phase stability, much like the displaced Sagnac interferometer described in the main part of the paper. The arrow on each walk-off crystal indicates the direction of its optic axis.}
\label{LinearInterferometer} 
\end{figure}

Alternatively, PBSs in Fig. \ref{MZPBS} can be implemented using two birefringent walk-off crystals (e.g. calcite) in a linear configuration\cite{OBrien03,Evans10}. The VPPBS based on the resulting interferometer is shown in Fig. \ref{LinearInterferometer}. In this design, the separation between optical paths is small and all paths pass through the same crystals. Consequently, much like the displaced Sagnac interferometer, this linear interferometer exhibits an inherent phase stability.

\section{Phase introduced by a tilted optical plate} \label{TiltMedia}

\begin{figure}[htbp]
\centering \includegraphics[width=\linewidth]{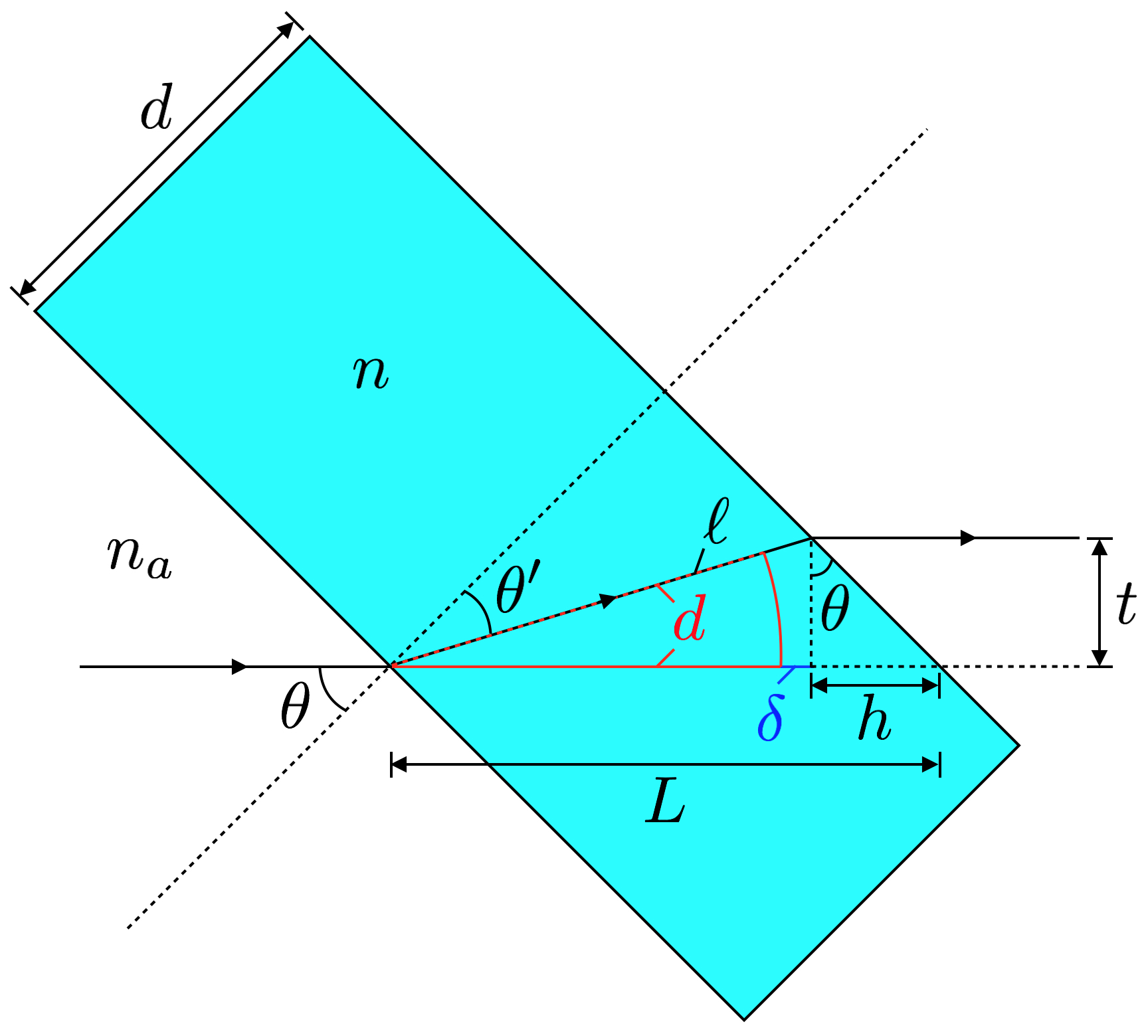} 
\caption{Optical medium of thickness $d$ and refractive index $n$ tilted
by an angle $\theta$. Such an angle has been exaggerated to introduce
all the important quantities in the calculation of the relative phase
$\phi$ in Eq. (\ref{phi}).}
\label{RelPhase} 
\end{figure}

In Sec. \ref{Theory}, the relative phase $\phi$
in Eq. (\ref{phi}) is obtained as follows. Consider an optical medium
of thickness $d$ and refractive index $n$ that is tilted by an angle
$\theta$, as shown in Fig. \ref{RelPhase}.
As mentioned in Sec. \ref{Theory}, when the optical medium is perpendicular
to the input beam the relative phase between the optical paths in
a Mach-Zehnder interferometer is given by 
\begin{equation}
\phi=\frac{2\pi d}{\lambda}(n-n_{a}),
\end{equation}
with $\lambda$ the light wavelength and $n_{a}$ the refractive index
of air. When the optical medium is tilted, the relative phase between
the two optical paths corresponds to 
\begin{equation}
\phi=\frac{2\pi\ell}{\lambda}n-\frac{2\pi(d+\delta)}{\lambda}n_{a},
\label{phiddelta}
\end{equation}
where $\ell$ is the length that light travels through the optical
medium, and $\delta$ is a small length that, together with $d$,
defines the longitudinal component of $\ell$. One can understand
Eq. (\ref{phiddelta}) considering two optical media, the first one
of thickness $\ell$ and refractive index $n$ in one arm of the interferometer
and the second one of thickness $d+\delta$ and refractive index $n_{a}$
(``made'' of air) in the other arm.

According to Fig. \ref{RelPhase}, $\ell$ is equal to 
\begin{equation}
\ell=\frac{d}{\cos\theta'},\label{ell}
\end{equation}
where $\theta'$ is the angle of refraction given by Snell's law in
Eq. (\ref{Snell}). In terms of $\ell$, the transverse separation
$t$ between the input and output beams is equal to 
\begin{equation}
t=\ell\sin(\theta-\theta')=\frac{d}{\cos\theta'}\sin(\theta-\theta'),
\end{equation}
allowing to find an expression for $h$ in Fig. \ref{RelPhase}, 
\begin{equation}
h=t\tan\theta=\frac{d}{\cos\theta'}\sin(\theta-\theta')\tan\theta.\label{h}
\end{equation}

On the other hand, the quantity $L$, defined as the total length
that light would travel through if there were no optical medium, is
\begin{equation}
L=\frac{d}{\cos\theta}.\label{L}
\end{equation}

However, according to Fig. \ref{RelPhase}, $d+\delta$ is equal to
$L-h$. Thus the relative phase in Eq. (\ref{phiddelta}) becomes
\begin{equation}
\phi=\frac{2\pi\ell}{\lambda}n-\frac{2\pi(L-h)}{\lambda}n_{a}.
\end{equation}

Substituting Eqs. (\ref{ell}), (\ref{h}) and (\ref{L}) into the last
expression, $\phi$ reduces to 
\begin{multline}
\phi=\frac{2\pi d}{\lambda}\\
\times\left[\frac{n}{\cos\theta'}-n_{a}\left(\frac{1}{\cos\theta}-\frac{\sin(\theta-\theta')\tan\theta}{\cos\theta'}\right)\right],
\end{multline}
which can be simplified to finally obtain Eq. (\ref{phi}) in Sec.
\ref{Theory}, 
\begin{equation}
\phi=\frac{2\pi d}{\lambda}\left[\frac{n}{\cos\theta'}-n_{a}\left(\cos\theta+\sin\theta\tan\theta'\right)\right].
\end{equation}
\end{document}